\documentclass[12pt]{article}
\usepackage{bbm}
\usepackage{graphicx}
\usepackage{cite}
\usepackage{amsmath}
\usepackage{amssymb}
\begin{document}
\title{\normalsize \hfill UWThPh-2005-3 \\
\normalsize \hfill SISSA-11-2005-EP
\\*[5mm] \LARGE
Unitarity triangle test of the extra factor of two\\ 
in particle oscillation phases} 
\author{
Samoil M. Bilenky$\,^{a\,b\,}$\thanks{E-mail: bilenky@sissa.it}\,,
\setcounter{footnote}{2}
Walter Grimus$\,^c\,$\thanks{E-mail: walter.grimus@univie.ac.at}\,,
\and
Thomas Schwetz$\,^a\,$\thanks{E-mail: schwetz@sissa.it} \\[4mm]
\small $^a$ Scuola Internazionale Superiore di Studi Avanzati \\
\small Via Beirut 2-4, I--34014 Trieste, Italy
\\*[3mm]
\small $^b$ Joint Institute for Nuclear Research, R--141980 Dubna, Russia 
\\*[3mm]
\small $^c$ Institut f\"ur Theoretische Physik, Universit\"at Wien \\
\small Boltzmanngasse 5, A--1090 Wien, Austria}

\date{February 18, 2005}

\maketitle

\begin{abstract}
There are claims in the literature that in neutrino oscillations and
oscillations of neutral kaons and $B$-mesons the oscillation phase
differs from the standard one by a factor of two. We reconsider the
arguments leading to this extra factor and investigate, in particular,
the non-relativistic regime. We actually find that the very same
arguments lead to an ambiguous phase and that the extra factor of two
is a special case.  We demonstrate that the unitarity triangle (UT)
fit in the Standard Model with three families is a suitable means to
discriminate between the standard oscillation phase and the phase with
an extra factor of two. If $K_L - K_S$ and $B_{dH} - B_{dL}$ mass
differences are extracted from the $K^0 - \bar K^0$ and $B_d^0 - \bar
B_d^0$ data, respectively, with the extra factor of two in the
oscillation phases, then the UT fit becomes significantly worse in
comparison with the standard fit and the extra factor of two is
disfavoured by the existing data at the level of more than $3\,
\sigma$.
\end{abstract}

\newpage

\section{Introduction}

Compelling evidence in favour of neutrino oscillations obtained in
recent years in the Super-Kamiokande \cite{SK-atm-1998,SK-solar}, SNO
\cite{SNO}, KamLAND \cite{KamLAND}, K2K \cite{K2K} and other neutrino
experiments (see e.g.~\cite{maltoni} and references therein) is a
major breakthrough in the search for physics beyond the Standard
Model.
All existing neutrino oscillation data with the exception of the LSND
data \cite{LSND}\footnote{The result of the LSND experiment is planned
to be checked by the MiniBooNE experiment \cite{miniboone} which is
currently taking data.}  are well described if we assume
three-neutrino mixing.  Defining $\Delta m^2_{jk}= m_j^2 - m_k^2$,
where the $m_j$ are the neutrino masses, the best fit values
\begin{equation}
\Delta m^{2}_{21} = 7.9  \times 10^{-5}\:\mathrm{eV}^{2}
\quad \mbox{and} \quad 
\left| \Delta m^{2}_{32} \right| = 2.4 \times 10^{-3}\:\mathrm{eV}^{2},
\label{1}
\end{equation}
were found for the solar \cite{KamLAND} and atmospheric neutrino
neutrino mass-squared differences \cite{SK-atm}, respectively.

These values of the neutrino mass-squared differences were obtained
from neutrino oscillation data under the assumption that the neutrino
transition and survival probabilities have the standard form (see
e.g.\ the reviews in Ref.~\cite{reviews}).  Neutrino oscillations are
due to the interference of the amplitudes of the propagation of
neutrinos with different masses and the standard phase differences are
given by the expression
\begin{equation}
\Delta\varphi_{jk}=\frac{\Delta m^2_{jk} L}{2E}.
\label{2}
\end{equation}
Here $E$ is the neutrino energy and $L$ is the 
distance between neutrino production and neutrino 
interaction points. The theory of neutrino oscillations has a long
history starting with the paper of Gribov and Pontecorvo \cite{gribov}
(for other early papers see \cite{pontecorvo,fritzsch}, for
historical overviews see \cite{nuhistory}). 
There is also a rich literature on more elaborate derivations of
neutrino transition and survival probabilities based on quantum
mechanics and quantum field theory (for a choice of these papers see
\cite{kayser,okun,nussinov,giunti93,rich,stockinger,stodolsky,%
cardall,kim,beuthe,lipkin04},  
more citations are found in the
reviews~\cite{zralek,giunti01,beuthe-review,giunti04}), 
which all result in the
standard oscillation phases of Eq.~(\ref{2}).

There exist, however, claims \cite{field} that the phase differences in
neutrino transition probabilities
differ from the standard ones by a factor of two and are equal to 
\begin{equation}
\overline{\Delta\varphi_{jk}}=\frac{\Delta m^2_{jk}L}{E}.
\label{3}
\end{equation}
Other authors \cite{deleo} claim that there is an ambiguity in the
oscillation phase. Theoretical discussions about the factor of two or
other factors in oscillation phases continue during many years---see
e.g.~\cite{kim,lipkin04,giunti04,tsukerman,lipkin} where these
additional factors have been refuted on theoretical grounds. Taking
into account the fundamental importance of the problem we believe that
it is worthwhile to think about possibilities to confront the
different oscillation phases to experimental data.

The same non-quantum-theoretical arguments which lead to an additional
factor of two in neutrino oscillation phases can be applied to the
oscillation phases in $M^0 \leftrightarrows \bar M^0$ oscillations of
neutral bosons $M^0 = K^0$, $B^0_d$, etc., as was demonstrated in
Ref.~\cite{lipkin}. A more complicated additional factor has been
obtained in Ref.~\cite{srivastava}, but was subsequently refuted in
Ref.~\cite{ancochea}.  Since in $M^0 \leftrightarrows \bar M^0$
oscillation experiments the mesons are often non-relativistic, the
relevant oscillation phase is
\begin{equation}\label{QMphase}
\Delta \varphi_\mathrm{QT} = \frac{\Delta m^2 L}{2p},
\end{equation}
where $p$ is the momentum of the neutral meson. In the
ultra-relativistic limit, Eq.~(\ref{QMphase}) coincides with
Eq.~(\ref{2}).  In the following we use the subscript QT for the
standard phase~(\ref{QMphase}), whereas phases different from the
standard phase are marked by a bar---see Eq.~(\ref{3}).

In recent years a remarkable progress in the measurement of $|V_{cb}|$,
$|V_{ub}|$ and other elements of the CKM matrix was reached (see
e.g.~\cite{CKM-group}). Another great achievement was the measurement
of the CP parameter $\sin2\beta$ with an accuracy of about 5\% in the 
BaBar \cite{babar} and Belle \cite{belle} experiments at asymmetric
B-factories. 
This allowed to perform a new check of the Standard Model based on the
test of the unitarity of the CKM mixing matrix, 
the so-called unitarity triangle test of the SM. 
It was shown \cite{buras1,buras2,silva,UT2000,UT2005} that the SM with three
families of quarks is in an good agreement with existing data, which
include the data on the measurements of the effects of CP violation. 
In the unitarity triangle (UT) test the experimental values of the
$K_L-K_S$ mass difference $\Delta m_K$ and the $B_{dH}-B_{dL}$ 
mass difference $\Delta m_{B_d}$ are used. 
The values of $\Delta m_{K}$ and $\Delta m_{B_d}$ were obtained from
an analysis of the experimental data based on the standard transition
probabilities with the standard oscillation phase~(\ref{QMphase}).

In this paper we will present the result of the UT
test under the assumption that oscillation phases in 
$K^0 \leftrightarrows \bar K^0$ and $B^0_d \leftrightarrows \bar B^0_d$ 
oscillations differ from
the standard ones by the above factor of two. 
We will show that such an assumption is disfavoured by the
existing data at the level of more than $3\,\sigma$.

The plan of the paper is as follows. 
In Section~\ref{notorious} we will discuss in some detail how this
notorious factor of two in the oscillation phase  
appears. Considerations how to confront the factor of two with
experiment are found in Section~\ref{confronting}.
Section~\ref{fit} contains our UT fit with and without
the factor of two. Our conclusions are presented in
Section~\ref{concl}. The technical details of the UT fit are deferred
to an appendix.

\section{The notorious factor of two}
\label{notorious}

\subsection{Notation}

For simplicity we consider oscillations between only two states. 
Thus we have two different masses $m_j$ ($j=1,2$). We adopt the
convention $m_1 < m_2$. For each mass eigenstate 
the relevant phase is
\begin{equation}\label{phi}
\varphi_j = E_j t - p_j L,
\end{equation}
where $E_j = \sqrt{p_j^2 + m_j^2}$ and $p_j$ are energy and momentum,
respectively.
Though there are some arguments that in particle oscillations mass
eigenstates with the same energies are coherent 
\cite{stockinger,stodolsky,lipkin04,lipkin},
we want to be general and assume neither equal energies nor equal momenta.

It is useful to define quantities $\Delta p$ and $\Delta m$ via 
\begin{equation}\label{averagequantities}
p_{1,2} = p \mp \frac{1}{2} \Delta p, \quad 
m_{1,2} = m \mp \frac{1}{2} \Delta m,
\end{equation}
where $p$ and $m$ denote average momentum and mass, respectively.
Defining $\Delta m^2 = m_2^2 - m_1^2$ and $\Delta m = m_2 - m_1$, we
have the relation
\begin{equation}
\Delta m^2 = 2m \Delta m.
\end{equation}
In the following we will use the approximations
\begin{equation}\label{approximations}
p \gg |\Delta p| \quad \mbox{with} \quad \Delta p = a \Delta m.
\end{equation}
The dimensionless constant $a$ is zero for $p_1 = p_2$. In general it
will be of order one or even larger. In the non-relativistic case one
can have $a \sim m/p$. 
The first relation of Eq.~(\ref{approximations})
excludes particles which are nearly at rest; such a situation is not
contained in our discussion. Consequently, we do not allow for 
$p \ll m$ or $a \gg 1$. 
However, we will take care that
all our considerations hold also in the moderately non-relativistic limit.
The second relation in Eq.~(\ref{approximations}) states our coherence
assumption: mass eigenstates with momenta which differ more 
than the mass difference can be coherent. 
Note that with Eq.~(\ref{approximations}) we have
\begin{equation}
p \gg \Delta m.
\end{equation}
In the following we will need 
\begin{equation}\label{diffE}
\Delta E \equiv E_2 - E_1 = 
\frac{1}{E} \left( m \Delta m + p \Delta p \right) =
\frac{\Delta m^2}{2E} + \frac{p \Delta p}{E} 
\quad \mbox{with} \quad E = \frac{1}{2} \left( E_1 + E_2 \right).
\end{equation}

\subsection{``Derivation'' of extra factors in oscillation phases}

Particle oscillation phases different from that of Eq.~(\ref{QMphase})
have been found for instance in Refs.~\cite{field,srivastava}, and an
ambiguity of a factor of two in the oscillation phase has been
diagnosed in Ref.~\cite{deleo}. It was stressed first in
Ref.~\cite{lipkin} and then in Refs.~\cite{giunti04,tsukerman} that in
essence the discrepancy to the standard result~(\ref{QMphase}) is due
to the assumption that the two mass eigenstates are detected at the
same space point but at different times
\begin{equation}\label{wrong}
t_j = L/v_j = LE_j/p_j.
\end{equation}
For each mass eigenstate, the corresponding time $t_j$ is inserted
into the phase (\ref{phi}). The motivation 
for this is that particles with different masses move with 
different velocities $v_j$. This picture mixes 
quantum-theoretical and classical considerations 
in an ad hoc fashion and leads to the
conclusion that particle phases taken at \emph{different times},
though at the same space point, produce the interference, which is in
contradiction to the rules of quantum theory. 

Eq.~(\ref{wrong}) gives the phase 
\begin{equation}
\overline{\varphi}_j = E_j t_j - p_j L = 
\frac{E_j^2 L}{p_j} - p_j L = \frac{m_j^2 L}{p_j} 
\end{equation}
and, therefore, the phase difference 
\begin{equation}\label{wrongdiff}
\overline{\Delta \varphi} = 
\frac{m_2^2 L}{p_2} - \frac{m_1^2 L}{p_1}.
\end{equation}
Then, using only $\Delta p \ll p$, we obtain
\begin{equation}\label{nonQM}
\overline{\Delta \varphi} \simeq 
2\, \Delta \varphi_\mathrm{QT} - 
\frac{\left( m_1^2 + m_2^2 \right) \Delta p\, L}{2\, p^2}.
\end{equation}
As seen from this equation, $\overline{\Delta \varphi}$ differs from
$\Delta \varphi_\mathrm{QT}$ not only by a factor of two, but also by
an additional term which contains the \emph{arbitrary}
quantity\footnote{In principle, one should be able to determine an
upper limit on $\Delta p$ from the widths of the wave packets of the
particles participating in the neutrino, $K^0$, $B_d^0$, etc.\
production and detection processes \cite{giunti93,stockinger,beuthe}.}
$\Delta p$. In the ultra-relativistic case, which always applies to
neutrinos but also to $M^0 \leftrightarrows \bar M^0$ oscillations when
their energy is high enough, the additional term is negligible and we
have the ultra-relativistic phase
\begin{equation}
\left( \overline{\Delta \varphi} \right)_\mathrm{UR} \simeq 
2\, \Delta \varphi_\mathrm{QT}.
\end{equation}
For oscillations of non-relativistic neutral flavoured mesons, the
additional term can not only be comparable with the first term but
could even dominate in Eq.~(\ref{nonQM}).  Since $\Delta p$ is
arbitrary, we come to the conclusion that, for oscillations of
non-relativistic particles, Eq.~(\ref{wrong}) leads to an
arbitrary---and thus unphysical---oscillation phase.

In order to illustrate the latter point, let us consider the two 
extreme cases of equal momenta and equal energies. 
In the first case with
$\Delta p = 0$, Eq.~(\ref{nonQM}) gives
\begin{equation}\label{factor2}
\overline{\Delta \varphi} = 
\frac{\Delta m^2 L}{p} = \frac{2m \Delta m L}{p}.
\end{equation}
Clearly, we have again the notorious factor of two, in comparison with
the quantum-theoreti\-cal result.
On the other hand, equal energies correspond to 
$\Delta p = -\Delta m^2/(2p)$ (see Eq.~(\ref{diffE})) 
and with Eq.~(\ref{nonQM}) the result is
\begin{equation}\label{factor2'}
\overline{\Delta \varphi} = 
\frac{\Delta m^2 L}{p} \left( 1 + \frac{m^2}{2p^2} \right).
\end{equation}
This oscillation phase, which is similar to the one advocated in
Ref.~\cite{srivastava}, agrees with Eq.~(\ref{factor2}) only in the
ultra-relativistic limit.

\subsection{The quantum-theoretical oscillation phase}

Although it has been stressed many times (see
e.g.\ Ref.~\cite{giunti01}) that the quantum-theoretical oscillation
phase does \emph{not} suffer from any ambiguity, it is instructive to
repeat the derivation of this fact here, in order to compare with the
derivation of Eq.~(\ref{nonQM}). Quantum theory requires the two
phases~(\ref{phi}) to be taken at the \emph{same space-time
point}. Therefore, we have
\begin{equation}
\Delta \varphi_\mathrm{QT} = \Delta E\, T - \Delta p L,
\end{equation}
where $T$ characterizes the time when the interference takes
place. 
Then, with $T = LE/p$ we obtain the quantum-theoretical result
\begin{equation}\label{phiQM}
\Delta \varphi_\mathrm{QT} = 
\left( \frac{\Delta m^2}{2E} + \frac{p \Delta p}{E} \right)
\frac{EL}{p} - \Delta p L = \frac{\Delta m^2L}{2p} = 
\frac{m\Delta mL}{p},
\end{equation}
where the arbitrary quantity $\Delta p$ 
has dropped out.\footnote{It is reasonable to assume 
that $T$ is $L/v_1$ or $L/v_2$
or some average of these two expressions. 
What one takes precisely as $T$ is
irrelevant, because all these 
possibilities differ only in terms suppressed by $\Delta m$ and 
$\Delta p$. Since $\Delta E$ is already small in that sense 
(see Eq.~(\ref{diffE})) and the first order in 
$\Delta m$ and $\Delta p$ is sufficient, we take the velocity $p/E$.}
For $M^0 \leftrightarrows \bar M^0$ oscillations, 
the phase~(\ref{phiQM}) can also be written in the familiar form 
$\Delta m\, \tau$, where $\tau$ is the eigentime of the particle for 
covering a distance $L$.

We want to emphasize that a more complete understanding of the
oscillation phase needs a full quantum-mechanical or quantum
field-theoretical approach. All such treatments (see for instance the
reviews~\cite{zralek,beuthe-review,giunti04} and 
references therein) consistently give the result of
Eq.~(\ref{phiQM}). In approaches not guided by quantum mechanics or
quantum field theory the conversion of time into a distance is always
the subtle point \cite{lipkin,ancochea}. In all present experiments, 
oscillations are treated as phenomena in space. If eigentimes are used
for the evaluation of data, then distances are converted into times
(see e.g.~\cite{babar,belle,hummel}).

\section{Confronting non-quantum-theoretical phases with experiment} 
\label{confronting}

Since we have seen that the derivation of phase~(\ref{nonQM}) does not
conform to the rules of quantum theory whereas Eq.~(\ref{QMphase})
does, then one could ask the question why consider the
phase~(\ref{nonQM}) at all. From our point of view, the reason for
this is twofold:
\begin{itemize}
\item
On the one hand, there is the subtlety that the time difference
$\Delta t = \left| t_2 - t_1 \right|$ (see Eq.~(\ref{wrong})), which
is the culprit of the discrepancy with the quantum-theoretical result,
is immeasurably small.
\item
On the other hand, as we will show, the phases~(\ref{factor2}) and
(\ref{factor2'}) can actually be tested experimentally.
\end{itemize}

The time difference can be expressed as
\begin{equation}\label{timediff}
\Delta t \simeq 
\frac{L}{2pE} 
\left| \Delta m^2 - 
\left( m_1^2 + m_2^2 \right) \frac{\Delta p}{p} \right|.
\end{equation}
To get a feeling for the size of $\Delta t$, we take the $K^0 \bar
K^0$ system with $\Delta m_K \simeq 3.48 \times 10^{-12}$~MeV and use
for example $L = 1$~m, $p = 1$~GeV and $\Delta p = 0$. Then we find
$\Delta t \sim 5 \times10^{-24}$~sec, which is indeed far beyond
measurability.

As for an experimental test of the phase~(\ref{factor2'}) we consider
two different measurements of the $K_L-K_S$ mass difference. Since
this phase has an additional dependence on the momentum, it is useful
to compare two measurements which have different average kaon
momenta. The CPLEAR experiment has measured \cite{CPLEAR}  
$\Delta m_K = (5295 \pm 20 \pm 3) \times 10^6\; 
\hbar \mathrm{s}^{-1}$. In that experiment kaons are produced in the
reaction $p \bar p \to K^+ \pi^- {\bar K}^0$ and the charged-conjugate
reaction, with $p \bar p$ annihilation at rest. Thus the kaons are
non-relativistic. In the KTeV experiment
the kaons are in the ultra-relativistic regime; this experiment has
obtained \cite{KTeV}
$\Delta m_K = (5261 \pm 15) \times 10^6\; 
\hbar \mathrm{s}^{-1}$. 
According to Eq.~(\ref{factor2'}) the mass differences extracted in these
experiments should be different and related by
\begin{equation}\label{ratio}
\frac{\left( \Delta m_K \right)_\mathrm{CPLEAR}}%
{\left( \Delta m_K \right)_\mathrm{KTeV}} =
1 + \frac{m_{K^0}^2}{2\,p_{K^0}^2} \geq 
1 + \frac{m_{K^0}^2}{2\,p_{K^0\,\mathrm{max}}^2},
\end{equation}
where $p_{K^0}$ is the (average) neutral-kaon momentum in the CPLEAR
experiment.\footnote{If Eq.~(\ref{factor2'}) were correct, there
should also be a dependence of the extracted mass difference on
$p_{K^0}$.}  One can show that the maximal energy of the neutral kaon
in the CPLEAR reaction is given by
\begin{equation}\label{Emax}
E_{K^0\,\mathrm{max}} = 
\frac{4\, m_p^2 - m_\pi^2 - 2\, m_\pi m_K}{4\, m_p},
\end{equation}
where $m_p$, $m_\pi$ and $m_K$ are proton, pion and kaon mass,
respectively. For our purpose the distinction between the mass values
of the charged and neutral kaon masses is irrelevant.  With the
numbers above for the mass differences obtained by the CPLEAR and KTeV
experiments, we use the law of propagation of errors to compute the
value $1.006 \pm 0.005$ for the ratio on the left-hand side of
Eq.~(\ref{ratio}). We insert the values of the particle masses into
Eq.~(\ref{Emax}) and calculate $p_{K^0\,\mathrm{max}}$; then we arrive
at 1.22 for the right-hand side of Eq.~(\ref{ratio}), which is about
40 standard deviations larger than the ratio of $K_L - K_S$ mass
differences.  Consequently, we conclude that the
phase~(\ref{factor2'}) is in contradiction to the results of the
CPLEAR and KTeV experiments.

The phase~(\ref{factor2}) which contains the notorious factor of two
needs a different approach; in the next section we will use the fit to
the unitarity triangle constructed from the CKM matrix to show that
this factor of two is experimentally strongly disfavoured.  For the
idea to compare the $\Delta m^2$ result of the solar neutrino
experiments with that of the KamLAND experiment see
Ref.~\cite{smirnov}.

\section{The unitarity triangle fit}
\label{fit}

\subsection{Description of the unitarity triangle analysis}
\label{sec:fit-description}

Following the traditional way, the unitarity triangle (UT) 
is given by the three points $A
= (\bar\rho,\bar\eta)$, $B = (1,0)$, $C = (0,0)$ in the plane of the
parameters $\bar\rho$ and $\bar\eta$, which are defined by 
\begin{equation}
\bar\rho = \rho \left( 1 - \frac{\lambda^2}{2} \right) \,,\quad
\bar\eta = \eta \left( 1 - \frac{\lambda^2}{2} \right) \,,
\end{equation}
where $\lambda,\rho,\eta$ are the Wolfenstein parameters of the CKM
matrix. Pedagogical introductions to the UT can be found e.g.\ in
Refs.~\cite{buras1,buras2,silva}.  
Our numerical analysis is based on the input data
as given in Tab.~1 of Ref.~\cite{UT2005}, and we use the following
constraints to determine the point $A = (\bar\rho,\bar\eta)$:
\begin{itemize}
\item
The measured value of $\varepsilon_K = (2.280\pm0.013)\times
10^{-3}$. The theoretical prediction for this quantity, which is a
measure for CP violation in $K^0 - \bar K^0$ mixing, is given
by\footnote{For the sake of brevity we drop the phase factor 
$\exp (i\pi/4)$ in $\varepsilon_K$, since it plays no role in the following.}
\begin{equation}\label{eq:epsilon}
\varepsilon_K = \frac{\hat B_K \, C}{\Delta m_K} \, 
\bar\eta \, \left[ (1-\bar\rho) \, D - E \right] \,,
\end{equation}
where $\Delta m_K$ is the $K_L-K_S$ mass difference and $\hat B_K, C,
D, E$ are numbers which have to be calculated and/or depend on
measured quantities such as $\lambda,\, m_t,\, m_c,\, |V_{cb}|$ 
(see e.g.\ Ref.~\cite{buras2} for precise definitions).
\item
The experimental determination of $|V_{ub}/V_{cb}|$.  This ratio is
connected to $\bar\rho,\bar\eta$ by
\begin{equation}
\sqrt{\bar\eta^2 + \bar\rho^2} = 
\left( \frac{1}{\lambda} - \frac{\lambda}{2} \right)
\left| \frac{V_{ub}}{V_{cb}} \right| \,.
\end{equation}
\item
The measurement of the $B_{dH} - B_{dL}$ mass difference
\begin{equation}\label{eq:DmBd}
\Delta m_{B_d} = 0.502\pm0.006\,\hbar\mathrm{ps}^{-1} \,.
\end{equation}
The theoretical prediction for the square root of $\Delta m_{B_d}$
as a function of $\bar\rho,\bar\eta$ is given by
\begin{equation}\label{eq:DmBdconstraint}
\sqrt{\Delta m_{B_d}} = F \, |V_{cb}| \, \lambda
\sqrt{\bar\eta^2 + (1 - \bar\rho)^2} \,,
\end{equation}
where $F$ is a constant depending on $m_t$ and other quantities
subject to theoretical uncertainties (see e.g.\ Ref.~\cite{buras2}).
\item
In addition we use direct information on the angles of the unitarity
triangle $\alpha,\beta,\gamma$. The angle $\beta$ has been measured
at BaBar and Belle, and we use the value $\sin2\beta =
0.726\pm0.037$.  For $\gamma$ we use the value $(59.1\pm16.7)^\circ$
(see Ref.~\cite{UT2005} for details), whereas for $\alpha$ we use the
likelihood function extracted from Fig.~10 of Ref.~\cite{UT2005} to
take into account the two allowed regions for $\alpha$ around
$107^\circ$ and $176^\circ$.
\end{itemize}
We do not use the constraint from $\Delta m_{B_s}$ which usually is
included in UT fits. The reason is that at present only a lower bound
exists on $\Delta m_{B_s}$, and therefore no further constraint is
obtained for the oscillation phase with the extra factor of two.
However, we remark that once an upper bound on $\Delta m_{B_s}$ will
have been established in the future, this will provide additional
information on the oscillation phase.

The fit is performed by constructing a $\chi^2$-function
$\chi^2(\bar\rho,\bar\eta)$ from these observables, including
experimental as well as theoretical errors. Technical details on our
analysis are given in Appendix~\ref{appendix}.

\subsection{Results of the UT analysis}

\begin{figure}
\centering
\includegraphics[width=0.7\textwidth]{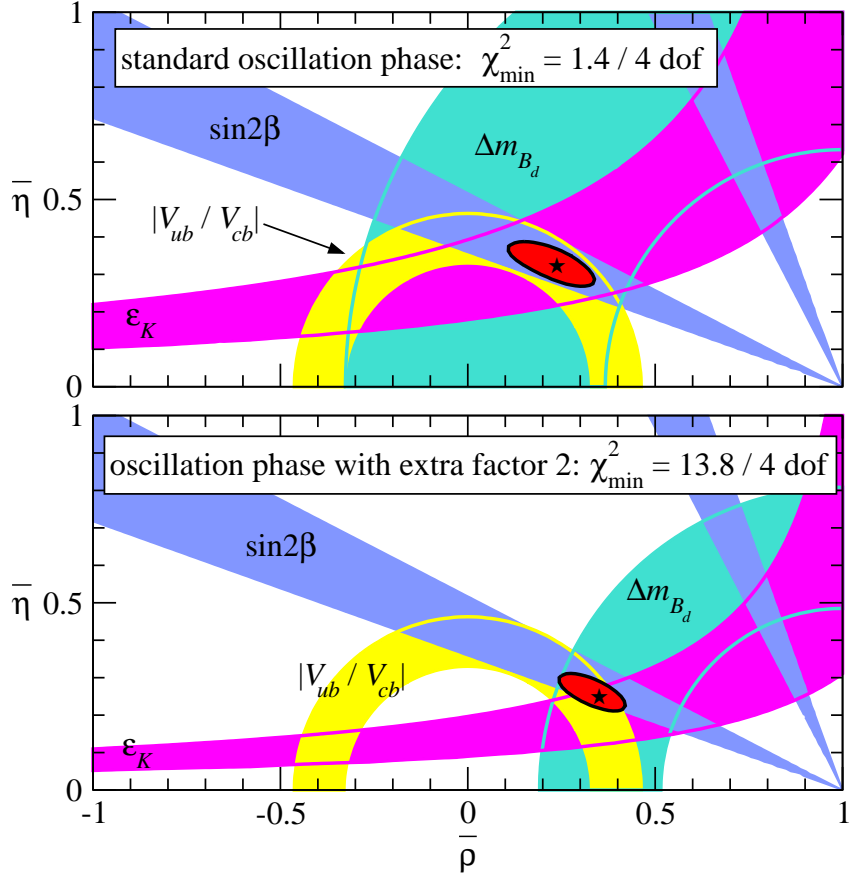}
\caption{Unitarity triangle fit with $\Delta m_K$ and $\Delta m_{B_d}$
obtained from the standard oscillation phase (upper panel) and the
oscillation phase with the extra factor of two (lower panel). The shaded
regions correspond to the 95\% CL regions (2 dof) obtained from the
constraints given by $\varepsilon_K, |V_{ub}/V_{cb}|, \Delta m_{B_d}$ and
$\sin2\beta$. In addition, constraints from the measurement of the
angles $\alpha$ and $\gamma$ are used in the fit (not shown in the
figure). The ellipses correspond to the 95\% CL regions from all data
combined, and the stars mark the best fit points.}
\label{fig:UT}
\end{figure}

The result of the standard UT fit is shown in the upper panel of
Fig.~\ref{fig:UT}. It is in good agreement with the results of various groups
performing this analysis, compare e.g.\ Refs.~\cite{CKM-group,UT2005,RPP}. We
show the allowed regions in the plane of $\bar\rho$ and $\bar\eta$ at 95\%~CL
for the individual constraints from $\varepsilon_K, |V_{ub}/V_{cb}|, \Delta
m_{B_d}, \sin 2\beta$, as well as the combined analysis including in addition
the information on $\alpha$ and $\gamma$. The 95\%~CL regions are obtained
within the Gaussian approximation for 2 degrees of freedom (dof), i.e.\ they
are given by contours of $\Delta\chi^2 = 5.99$.  For the best fit point of the
combined analysis we obtain $\bar\rho = 0.237, \bar\eta = 0.325$ with the
95\%~CL allowed region shown as the ellipse in Fig.~\ref{fig:UT}. Assuming
that the $\chi^2$-minimum has as a $\chi^2$-distribution with $(6-2)$ dof our
value of $\chi^2_\mathrm{min} = 1.4$ implies the excellent goodness of fit of
84\%.

Let us now discuss how an extra factor of two in the oscillation phase will
affect the UT fit. If such a factor is present the mass differences inferred
from particle--antiparticle oscillation experiments will be two times
smaller. Therefore, whenever in the UT analysis a mass difference
inferred from oscillations enters one has to use
\begin{equation}\label{eq:r}
\overline{\Delta m} = r\, \Delta m 
\end{equation}
with $r=1/2$, where $\Delta m$ is the value obtained with the standard
oscillation phase, i.e.\ this is the value which is given by the
Particle Data Group~\cite{RPP}. In the lower panel of
Fig.~\ref{fig:UT} we show the result of the UT fit by using the extra
factor of two in the oscillation phase. This factor affects two
observables relevant for the UT fit. 
\begin{enumerate}
\item
In the prediction for $\varepsilon_K$ shown in Eq.~(\ref{eq:epsilon}) the
experimental value for $\Delta m_K$ is used. Since this value is
obtained from $K^0 \leftrightarrows \bar K^0$ oscillations, 
$\Delta m_K$ has to be 
replaced by $\overline{\Delta m}_K$ if there is an extra factor of two in
the oscillation phase. This moves the hyperbola in the
($\bar\rho,\bar\eta$) plane from $\varepsilon_K$ to the right, as visible
in Fig.~\ref{fig:UT}.
\item
The experimental value for $\Delta m_{B_d}$ given in
Eq.~(\ref{eq:DmBd}) has to be replaced by $\overline{\Delta m}_{B_d}$,
which is a factor of two smaller. Therefore, from
Eq.~(\ref{eq:DmBdconstraint}) it is clear that the radius of the
circle in the ($\bar\rho,\bar\eta$) plane from $\Delta m_{B_d}$ is
reduced by a factor $\sqrt{2}$, as can be seen also in
Fig.~\ref{fig:UT}.
\end{enumerate}
The other constraints from $|V_{ub}/V_{cb}|, \sin2\beta, \alpha$ and $\gamma$
are obtained from particle decays without involving any oscillation effect,
and therefore they do not depend on the oscillation phase. One observes from
Fig.~\ref{fig:UT} that the agreement of the individual constraints gets
significantly worse using the extra factor of two. 
In particular, at 95\%~CL there
is only a very marginal overlap of the intersection of the allowed regions
from $|V_{ub}/V_{cb}|$ and $\sin2\beta$ with the one from $\varepsilon_K$. The
best fit point in the lower panel of Fig.~\ref{fig:UT} has
$\chi^2_\mathrm{min} = 13.8$, which implies a goodness of fit of only 0.8\%,
assuming a $\chi^2$-distribution for 4~dof.

\begin{figure}
\centering
\includegraphics[width=0.6\textwidth]{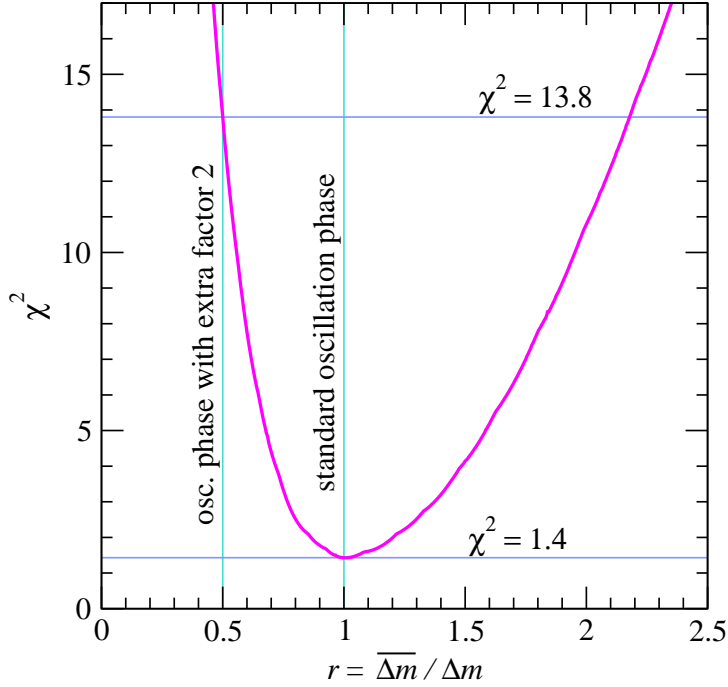}
\caption{$\chi^2$ of the unitarity triangle fit as a function of the
parameter $r$ defined in Eq.~(\ref{eq:r}). For fixed $r$ the $\chi^2$
is minimized with respect to $\bar\rho$ and $\bar\eta$.}
\label{fig:chisq}
\end{figure}

In Fig.~\ref{fig:chisq} we show the $\chi^2$ minimized with respect to
$\bar\rho$ and $\bar\eta$ as a function of the parameter $r$ given in
Eq.~(\ref{eq:r}). Hence, $r = 1$ corresponds to the standard
oscillation phase, and $r = 1/2$ corresponds to the extra factor
of two. From this figure one observes the remarkable feature that the best
fit point occurs exactly at $r=1$. In other words, even if 
the extra factor in the 
oscillation phase is treated as a free parameter to be determined by
the fit, the data prefer the standard oscillation phase. For the value
$r=1/2$ we obtain a $\Delta \chi^2 = 12.4$ with respect to the best
fit point, which corresponds to an exclusion at $3.5\sigma$ for
1~dof. We conclude that the extra factor of two in the oscillation
phase is strongly disfavoured by the UT fit.

\subsection{Robustness of the UT analysis}
\label{sec:robustness}

In this subsection we investigate the robustness of our conclusion with
respect to variations of the input data for the UT fit. To this aim we show in
Tab.~\ref{tab:variations} the results of our analysis by changing some of the
numbers entering the UT fit. The line ``standard analysis'' in the table
corresponds to the analysis described in the previous two subsections. In
particular, exactly the input data given in Tab.~1 of Ref.~\cite{UT2005} 
are used.

First we have investigated how our analysis depends on the value for
$|V_{ub}|$. We show the results of the fit by using only the value
from exclusive ($|V_{ub}|_\mathrm{(excl)}$) or inclusive
($|V_{ub}|_\mathrm{(incl)}$) decays, where the numbers are taken from
Ref.~\cite{UT2005}. Note that in our standard analysis both values are
taken into account, as described in Appendix~\ref{appendix}. We
observe from the numbers given in Tab.~\ref{tab:variations} that for
the relatively small value for $|V_{ub}|$ from exclusive measurements
the fit gets notably worse for $r=1/2$. In contrast, for the
relatively large value from inclusive measurements the fit gets worse
for the standard oscillation phase ($\chi^2_\mathrm{min} = 3.9$),
whereas for $r=1/2$ the fit improves with respect to the standard
analysis ($\chi^2_\mathrm{min} = 7.8$). The reason is that for large
values of $|V_{ub}|$ the radius of the circle in the
($\bar\rho,\bar\eta$) plane from $|V_{ub} / V_{cb}|$ becomes larger,
which worsens the fit for $r=1$, whereas for $r=1/2$ the agreement of
the individual allowed regions becomes better. Note however, that even
for $|V_{ub}|_\mathrm{(incl)}$ the goodness of fit for $r=1/2$ is only
1\%, and $r=1/2$ is disfavoured with respect to $r=1$ by 2$\sigma$.
We have also performed the analysis by using the (inclusive and
exclusive) averaged value $|V_{ub}|_\mathrm{(PDG)}$ obtained by the
PDG~\cite{RPP}. The fit using the extra factor of two is slightly improved
with respect to our standard analysis, however $r=1/2$ can still be
excluded at $3.2\sigma$.

\begin{table}
\centering
\begin{tabular}{|l|c|c|c|}
\hline\hline
 & $\chi^2_\mathrm{min} ( r =1)$
 & $\chi^2_\mathrm{min} ( r =1/2)$
 & number of $\sigma$ \\
\hline
standard analysis & 1.4 & 13.8 & 3.5 \\
\hline
$|V_{ub}|_\mathrm{(excl)} = (33.0 \pm 2.4 \pm 4.6)\times 10^{-4} $  
  & 2.9 & 17.6 & 3.8 \\
$|V_{ub}|_\mathrm{(incl)} = (47.0 \pm 4.4) \times 10^{-4}$ & 3.9 & 7.8 & 2.0 \\
$|V_{ub}|_\mathrm{(PDG)} = (36.7 \pm 4.7) \times 10^{-4}$ & 1.6 & 11.9 & 3.2 \\
\hline
$m_c = (1.2 \pm 0.2)$ GeV  & 1.4 & 11.9 & 3.2 \\
\hline
constraints on $\alpha,\gamma$ not used & 0.13 & 9.6 & 3.1\\
\hline\hline
\end{tabular}
\caption{The $\chi^2_\mathrm{min}$ for the standard oscillation phase ($r=1$)
  and for the oscillation phase with the extra factor of two ($r=1/2$) for
  variations of the input data (see text for details). The column ``number of
  $\sigma$'' gives the number of standard deviations with which $r=1/2$ is
  disfavoured with respect to $r=1$.}
\label{tab:variations}
\end{table}

Furthermore we have investigated how our result depends on the input
value for the charm quark mass $m_c$. The value $m_c = (1.2 \pm
0.2)$~GeV is adopted by the CKM-fitter group~\cite{CKM-group},
in contrast to the value $m_c = (1.3 \pm 0.1)$~GeV from the UTfit
Collaboration~\cite{UT2005} used in our standard analysis. The mild
improvement of the fit for $r=1/2$ comes mainly from the larger error
on $m_c$, which leads to a slightly larger allowed region from
$\varepsilon_K$. 

In the last line of Tab.~\ref{tab:variations} we have removed the
constraints for the angles $\alpha$ and $\gamma$ from the fit, i.e.\
we use only $\varepsilon_K, |V_{ub}/V_{cb}|, \Delta m_{B_d}, \sin
2\beta$. We observe that the direct constraints of $\alpha$ and
$\gamma$ contribute $4.2$ units of $\chi^2$ to the
$\chi^2_\mathrm{min}$ for $r=1/2$. However, also without the
constraints for $\alpha$ and $\gamma$ the extra factor of two in the
oscillation phase is excluded by more than $3\sigma$.

Finally let us comment on the the very small value of
$\chi^2_\mathrm{min} = 0.13$ (for 2~dof), which we obtain without the
constraints on $\alpha$ and $\gamma$ for the standard oscillation
phase. In fact, the $\chi^2$-minimum value of 1.4 in the standard
analysis comes mainly from $\alpha$. To include the information on
this angle we are using the likelihood function from Fig.~10 of
Ref.~\cite{UT2005} (see Appendix~\ref{appendix}), which has two maxima
around $107^\circ$ and $176^\circ$. The maximum at $176^\circ$ is
slightly preferred, whereas the UT fit requires the other maximum. The
very small $\chi^2$-minimum value obtained without using the
likelihood for $\alpha$ shows that the fit is dominated by rather
large theoretical errors. Therefore, $\chi^2$ is significantly lower
as expected just from statistics. The fact that even with these
assumptions on theoretical errors the $\chi^2$ is large for $r=1/2$
implies that the exclusion of the extra factor of two in the
oscillation phase is rather robust.

\section{Conclusions}
\label{concl}

In this paper we have reconsidered claims that the standard oscillation
phase~(\ref{phiQM}) has to be corrected by extra factors. We have 
focused on possible tests of these extra factors by using experimental
data. The usual starting point to derive these non-quantum-theoretical
expressions for the oscillation phase is Eq.~(\ref{wrong}), which  
says that mass eigenstates with different masses need different
times to reach the spatial point where the interference of the
amplitudes for the different mass eigenstates takes place. In this way
we have derived the phase $\overline{\Delta \varphi}$ of
Eq.~(\ref{nonQM}). The aim of our theoretical discussion was to
consider both neutrino oscillations and 
oscillations of neutral flavoured mesons. For $M^0 - \bar M^0$
oscillations, it was important to include the non-relativistic limit
in our phase considerations. 

We have obtained the following results:
\begin{enumerate}
\item
The non-quantum-theoretical phase 
$\overline{\Delta \varphi}$ of Eq.~(\ref{nonQM}) becomes ambiguous in
the non-relativistic case, because it contains a small but arbitrary
momentum difference $\Delta p$. We have stressed that in the correct
quantum-theoretical treatment, where the amplitudes interfere at the
\emph{same} time, this arbitrary term does \emph{not} show up.
\item
If we adjust $\Delta p$ in Eq.~(\ref{nonQM})
such that the mass eigenstates have the same
energy, then a momentum-dependent extra factor appears in 
$\overline{\Delta \varphi}$---see Eq.~(\ref{factor2'}). We have shown
that this extra momentum dependence is in disagreement with
measurements of the $K_L - K_S$ mass difference at different kaon
energies. 
\item
If $\Delta p = 0$, the notorious factor of two appears in 
$\overline{\Delta \varphi}$---see Eq.~(\ref{factor2}). 
We have demonstrated that using $K_L - K_S$ and $B_{dH} - B_{dL}$ 
mass differences extracted from the data with the extra factor of two in the 
$K^0 \leftrightarrows \bar K^0$ and 
$B_d^0 \leftrightarrows \bar B_d^0$ oscillation phases,
respectively, the unitarity triangle fit in the Standard Model becomes
significantly worse compared to the fit with the standard mass
differences. The phase with the extra factor of two is excluded at
more than three standard deviations with respect to the standard phase. 
\end{enumerate}
Concerning this last point, as an additional check, we have treated
the extra factor in the oscillation phase as a free parameter $r$ (see
Eq.~(\ref{eq:r})) and considered $\chi^2$ as a function of $r$. It is
remarkable that the minimum of $\chi^2$ occurs nearly precisely at
$r=1$, which corresponds to the standard oscillation phase.  This
result can be regarded as a successful test of quantum theory.  It is
likely that in the future, with accumulated data used in the unitarity
triangle fit, the exclusion of the extra factor of two will become
even more significant.

\vspace{5mm}

\noindent
\textbf{Acknowledgements:}
S.M.B.\ acknowledges the support by the 
Italian Program ``Rientro dei Cervelli''.
W.G.\ would like to thank S.T.\ Petcov for an invitation to SISSA, 
where part of this work was performed. He is also grateful to 
A.Yu.\ Smirnov for a useful discussion. 
T.S.\ is supported by a ``Marie Curie Intra-European Fellowship within
the 6th European Community Framework Programme.''

\begin{appendix}

\section{Details of our UT fit procedure}
\label{appendix}

The fit of the UT is performed by adopting the following $\chi^2$-function:
\begin{eqnarray}
\chi^2(\bar\rho,\bar\eta, \hat B_K, |V_{ub}|) &=&
\sum_{i,j} (x_i^\mathrm{exp} - x_i^\mathrm{pred}) S^{-1}_{ij}
(x_j^\mathrm{exp} - x_j^\mathrm{pred}) + 
\chi^2_\alpha \nonumber\\ 
&+&
\chi^2_\mathrm{syst}(\hat B_K) + \chi^2_\mathrm{syst}(|V_{ub}|) 
\label{eq:chisq}
\end{eqnarray}
The final $\chi^2$ is obtained by minimizing Eq.~(\ref{eq:chisq}) with
respect to $ \hat B_K$ and $|V_{ub}|$:
\begin{equation}
\chi^2(\bar\rho,\bar\eta) 
= 
\mathrm{Min} 
\left[\chi^2(\bar\rho,\bar\eta, \hat B_K, |V_{ub}|); \, \hat B_K, |V_{ub}|
\right] \,.
\end{equation}
In Eq.~(\ref{eq:chisq}) the indices $i,j$ run over $(\varepsilon_K,
|V_{ub}/V_{cb}|, \Delta m_{B_d}, \beta, \gamma)$ and $S_{ij}$ is the
covariance matrix of these observables containing the experimental as
well as theoretical uncertainties. It also takes into account correlations
between the various observables induced by the experimental errors of 
parameters such as $m_t,\lambda$ and $|V_{cb}|$, which are common to
more than one observable. 
The term $\chi^2_\alpha$ contains the information on the angle
$\alpha$, and is defined as $\chi^2_\alpha = -2 \ln
[\mathcal{L}(\alpha) / \mathrm{Max} \, \mathcal{L}(\alpha) ]$, where
$\mathcal{L}(\alpha)$ is the likelihood function for $\alpha$ read off
from Fig.~10 of Ref.~\cite{UT2005}.

For the treatment of theoretical uncertainties we follow the common
practice in UT fits to split the error into a Gaussian
part and into a ``flat'' part, which cannot be assigned a
probabilistic interpretation~\cite{CKM-group,UT2000,UT2005}. For the
parameter $\hat B_K$ relevant for $\varepsilon_K$ one has $\hat B_K =
0.86 \pm 0.06 \pm 0.14$, where the first error is Gaussian and the
second is ``flat''. To include both errors in our fit we construct a
likelihood function $\mathcal{L}(\hat B_k)$ by convoluting a Gaussian
distribution with width $0.06$ with a flat distribution which is
non-zero in the interval $[-0.14,+0.14]$ and zero outside. Then this
likelihood is converted into a $\chi^2$ by $\chi^2_\mathrm{syst}(\hat
B_K) = -2 \ln [ \mathcal{L}(\hat B_k) / \mathrm{Max} \mathcal{L}(\hat
B_k)]$ which is added to the total $\chi^2$ according to
Eq.~(\ref{eq:chisq}). The resulting $\chi^2$ is minimised for fixed
$\bar\rho$ and $\bar\eta$ with respect to $\hat B_K$.

The value of $|V_{ub}|$ can be obtained from exclusive and inclusive decays,
where the exclusive measurement suffers from theoretical uncertainties
characterized by a ``flat'' error (see e.g.\ Tab.~1 of Ref.~\cite{UT2005}). In
our standard analysis we include both values by constructing a likelihood
function $\mathcal{L}(|V_{ub}|) = \mathcal{L}_\mathrm{excl}(|V_{ub}|) \times
\mathcal{L}_\mathrm{incl}(|V_{ub}|)$, where
$\mathcal{L}_\mathrm{excl}(|V_{ub}|)$ is obtained similar as in the case of
$\hat B_K$ by folding a Gaussian and a flat distribution, whereas
$\mathcal{L}_\mathrm{incl}(|V_{ub}|)$ is just a Gaussian
distribution. Finally, the term $\chi^2_\mathrm{syst}(|V_{ub}|)$ in
Eq.~(\ref{eq:chisq}) is obtained by $\chi^2_\mathrm{syst}(|V_{ub}|) = -2 \ln [
\mathcal{L}(|V_{ub}|) / \mathrm{Max} \mathcal{L}(|V_{ub}|)]$. The dependence
of our results on the treatment of $|V_{ub}|$ is discussed in
Sec.~\ref{sec:robustness}.

\end{appendix}

\end{document}